\documentstyle[12pt]{article}

\title{p-adic probability prediction of correlations between
particles in the two-slit and neutron interferometry experiments}

\author{Andrei Khrennikov\\ Department of Mathematics, Statistics
and Computer Sciences,\\
University of V\"axj\"o, 35195, V\"axj\"o, Sweden}

\begin{document}
\maketitle

\footnote{\it On leave from Moscow Institute of Electronic Engineering.
These investigations were supported by
Alexander von Humboldt and Rikkyo University Fellowships.}

\begin{abstract} We start from Feynman`s idea to use negative
probabilities to describe the two slit experiment and other
quantum interfernce experiments. Formally by using  negative
probability distributions we can explain the results of the two
slit experiment on the basis of the pure corpuscular picture of
quantum mechnanics. However, negative probabilities are absurd
objects in the framework of the standard Kolmogorov theory of
probability. We present a large class of non-Kolmogorovean
probability models where negative probabilities are well defined
on the frequency basis. These are models with probabilities which
belong to the so-called field of $p$-adic numbers. However, these
models are characterized by correlations between trails.
Therefore, we predict correlations between particles in
interference experiments. In fact, our predictions are similar to
the predictions of the so-called nonergodic interpretation of
quantum mechanics, which was proposed by V. Buonomano. We propose
the concrete experiments (in particular, in the framework of the
neutron interferometry) to verify our predictions on the
correlations.
\end{abstract}

\section{Introduction}

It is well known (see,for example, R.Feynman [1]) that probabilistic
distributions which appear in the two slit experiment would not be
described by the usual theory of probability (based on the axiomatic of
A.N.Kolmogorov [2]). There were different attempts to propose new
probabilistic theories to describe this situation (see, [3]-[6]). Despite of
some of these theories were sufficiently fruitful from the formal
mathematical point of view, they could not explain the physical matter
of the two slit experiment, i.e., in particular, to answer to the
question: {\it does a quantum particle go through one fixed slit or through
both slits?}

In the present paper we also try to apply to the two slit experiment a
new probabilistic theory. This is so called a $p$-adic theory of
probability [7]. There probabilities might belong to a field of $p$-adic
numbers ${\bf Q}_p$ (as the field of real numbers ${\bf R},$ this field
is a completion of the field of rational numbers ${\bf Q}$ , see the next
section for the main properties of $p$-adic numbers).
\footnote{ First ${\bf Q}_p$-valued probabilities were used in so called
$p$-adic(non-Archimedean) physics (see, for example,
[8],[7] for these physical models).}
As probabilities do not belong to the segment [0,1] of the real line,
$p$-adic theory is a non-Kolmogorovean probabilistic model (see [9],[10]
on the
extensions of Kolmogorov`s axiomatic). But our approach differs very
much from previous non-Kolmogorovean descriptions of the two slit
experiment [3]-[6]. Our theory is a frequency theory, i.e. it is formulated,
not in the framework of a measure theory, but using a frequency
definition of probability. Hence we have  direct connection with
physical reality (using relative frequencies) and may test in experiments
some consequences of a $p$-adic probabilistic model for the two slit
experiment.

The main prediction of our $p$-adic theory of probability is that the
Kolmogorov (algorithmic) complexity of "random sequences" $\omega$
generated in the two slit experiment has the behaviour $K[(\omega)_n]=
\log_p n $ where $(\omega)_n= (\omega_1,...,\omega_n)$ is the initial segment
of $\omega$ of the length $n.$ On the other hand, (at least formally)
the Kolmogorov complexity of random sequences corresponding to independent
trials has the behaviour $K[(\omega)_n]= n.$

{\bf Conclusion.} {\it There exist correlations between quantum particles
in the two-slit (and other interference) experiments.}

This prediction can be verified experimentally.

As we hope that this paper should be interesting for
 experimentalists, we start directly from the experimental consequences
of the $p$-adic probability predictions. In particular, we discuss the
possible experiments to show the correlation between particles. We consider
also the framework of the neutron interferometry [11] and propose some concrete
experiments. In fact, these experiments are not so complicated (or expensive).
They could be realized on the standard equipment of the neutron interferometry.
If some of these experiments be successful, the pure particle picture of
quantum mechanics should be justified and interference ~phenomenon should be
regarded to the interaction between quantum particles and laboratory equipment.

The second part of the paper is devoted to the theoretical considerations.
In particular, this part contains all primary facts about $p$-adic numbers,
foundations of the frequency theory of probability (R.von Mises [12], 1919),
non-Kolmogorovean model with $p$-adic probabilities and the Kolmogorov
~algorithmic complexity.

In the theoretical ~background our main idea is the following:

{\it As geometry is not ~restricted to the Euclidean model, in the same
way probability is not restricted to the Kolmogorovean model.
As some physical phenomena cannot be described by the Euclidean geometry,
in the same way some physical phenomena  cannot be described by the Kolmogorov
probability.}

\section{Experimental consequences for the two slit experiment}

In the theoretical part of this paper we shall follow to the following
chain of ~considerations:

{\it experiment $\longrightarrow$ violations of the ordinary
probabilistic properties $\longrightarrow$ negative probabilities
solution $\longrightarrow$ $p$-adic frequency description of negative
probabilities in the two slit experiment $\longrightarrow$
Kolmogorov complexity of $p$-adic collectives $\longrightarrow$
correlations between trials in the two slit experiment.}

As a consequence, we have

{\bf Conclusion.} Trials in the two slit experiment are not independent.

We have to test our prediction in physical experiments. At the moment,
we do not know Where is an information about previous trails is
accumulated? There are three (less or more natural) possibilities:

(1) It is accumulated in the aperture. A new particle does not go
through the aperture independently with previous particles.

(2) Previous particles change a structure of the screen. The position
of a new particle on the screen depends on these previous changes.

(3) The source of particles accumulates an information about previous
particles.

It seems to be that (1) and (3) are the most important possibilities.

What kind of experiments may test these hypothesis?

To exclude the correlations in the source of particle,
we need a source of single particles which could not accumulate
the information on previous particles. In the ideal case, we
have to use a new source for a new particle.

To exclude the correlations due to (1) or (2),
we have to change both shields ( the shield with apertures and
the screen) after each single particle.
Hence, we should get only one point on every screen.
Finally we should construct the histogram of points using a large
statistical ensemble of screens with a single point on each of them.

We predict that {\it there should be no interference rings on this
histogram or at least the interference should be very weak.}

Then we may realize ~experiments to separate ~hypothesis (1)-(3). For
instance, we may change only screens  after every experiment with a
single particle.

These experiments seem to be very simple from the theoretical point of view.
However, the discussion with scientists working in the quantum measurements
showed that it should be technical problems to present a large
ensemble of the {\it identical} equipment for the two-slit experiment.

Therefore, we have to propose more real experiments to verify our predictions.
In the next section we shall discuss such experiments in the standard framework
of the neutron interferometry. Then we shall go back to the two-slit
framework and propose ~essentially new experiment to find the correlations
between quantum particles.

\section{Experimental consequences for the neutron interferometry}

 In fact, our predictions on the basis of the $p$-adic
probability theory coincide with predictions of the so-called nonergodic
interpretation of quantum mechanics, which was proposed by V. Buonomano
[13].\footnote{I should like to thank H. Rauch and J. Summhammer
who had pointed me to this connection.}
This interpretation uses the standard formalism of quantum theory, but
it associates the expression $< A \psi, \psi>,$ which denotes the
expectation value  of the observable $A$ of a system in state $\psi,$ with
the time overage, rather than the ensemble average. The
nonergodic interpretation of quantum mechanics was tested by few experiments
in the framework of the neutron interferometry [14] - [16].
However, the results of
these experiments seem to against the nonergodic interpretation of quantum
mechanics.
Therefore, they are more or less against our $p$-adic
probabilistic predictions.

In the framework of the neutron interferometry
we get the same predictions as for the two slit experiment:
{\it all the interference in interferometers is a result of the dependence
(correlations) of the detection events.} Therefore, I propose to reduce
these correlations which should result in reduced visibility of the ~interference
~fringes. In the ideal case, if there are no correlations at all, one should see
no interference.

In contrary all people working on neutron interferometry ~believe that detection
events are Poissonian distributed and therefore completely independent from each
other. For example, M. Zawisky [16]
 has done some experiments which tested the
distribution of the output beams. It was found no significant deviation from
the poissonian statistics within an accuracy of approximation $2\% .$
 Nevertheless
a visibility of $50\%$  of interference fringes in the outgoing beam was
detected.

However, although there were no evidence for deviations from Poissonian statistics,
it would be interesting to repeat the statistical analysis with more accuracy.
To do this, we need to know how much difference we could expect between
$p$-adic and Poissonian statistics. Of course, if the effect is too small,
then it probably will be difficult to measure it with neutron interferometry.

At the moment, we cannot estimate accuracy because the p-adic probabilistic
distribution is a type of hidden variables distribution. Our analysis implies
only that "this is a p-adic distribution", but we cannot describe the concrete
form of this distribution. For example, is it p-adic uniform distribution or
not?  Moreover, there is the parameter of the model: a prime number $p.$ There
can be 2-adic, 3-adic, 1997-adic distributions. Therefore it is not easy to
get some predictions on the accuracy.

The possibility to prove our theory would be to increase the independence
of all detection events and to measure the reduced visibility as a function of
this independency. The general proposal is to change the source, ~~interferometer
and phase shifter after each single event. But it seems to be impossible in
practice.

M. Zawisky proposed (in a private discussion)
a much cheaper and easy way to perform experiment:
If there are some memory effects in the system (neutron-interferometer,
phase-shifter, source), then these effects must be time independent, otherwise
the visibility would increase after ~each experiment. If we put different beam
attenuators in front of the interferometer and if we measure the visibility with
different input intensivities (but with same particles numbers therefore the
measurement time depends on the beam attenuation) one should see such time
dependent memory effects.

The most realistic is another experiment: to ~disturb the interferometer
at the beginning of each measurement cycle to guarantee zero visibility at
the starting point. In this experiment we may hope that we shall get
 non-Poissonian distribution.

Another idea is to destroy the time dependence by using the
time factor: to repeat experiment after sufficiently long time. Of course, the
main question is : What is a meaning of "long time"?
From the $p$-adic point of view we have the following exponential scale:
$t=p^n,$ i.e. $t_0=0$ (first experiment), $t_1=p$ (second),...., $t=p^n (
(n-1)$-experiment),...
Of course, there is still a problem of the parameter $p.$ The only possibility
is to start with $p=2.$ If the Poissonian structure is not destroyed
(i.e. time-scale is too small),  then to try $p=3,$ and so on.
If the interference picture is not totally disappeared, we could hope that
at least it will become more and more unsharp with increasing of $p.$

\section{Random two-slit experiment}

It seems to be that the following experiment should delete or at least
make weaker the memory effects in the shield with slits.

Let us consider the shield with a large number of slits: $S=
\{s_1,....,s_N\} , N=2^k.$
There is a device ${\cal D}$
which can open and close slits. Consider two generators
of (ordinary) pseudo-random numbers $\xi$ and $\eta.$
For example, these numbers can be chosen uniformly distributed on the set
$\{ 1,...,N \}.$
According
to the values of these random generators ${\cal D}$ open only two slits,
$s_{\xi},s_{\eta},$ for each particle generated by the source of single
particles ${\cal I}.$ Therefore, for any  quantum particle
registrated on the screen, we are in the framework of the two slit
experiment (of course, if $\xi=\eta$ for some trial, then we have
one slit experiment which is considered as a particular case of the two-slit
experiment). However, we need a separate screen for each particle registration.

Let us fix some configuration of the two-slits, $\alpha=(s_{i_1}, s_{i_2}).$
After a large number of experiments we collect all screens corresponding
to the experiments with $\xi=i_1$ and $\eta=i_2$ and construct the corresponding
distribution of points by the projection to the unique screen.

From the point of view of the ordinary quantum mechanics we should
get the standard ~interference picture which corresponds to the "pure
two-slit experiment" with the slits $\alpha=(s_{i_1}, s_{i_2}).$

However, our $p$-adic theory predicts that in the ideal case the interference
picture should disappear. The ideal case means that $N\to \infty.$ In any
case we predict that the interference picture should become weaker and
weaker with the increasing of $N.$

\section{Frequency theories of probability}

The $p$-adic frequency probability theory is a
natural extension of the von Mises theory  [12]
where probabilities were defined via a principle of statistical
stabilization of relative frequencies. According to this principle,
the statistical sample
\begin{equation}
\label{e1}
x=(x_1,x_2,...,x_n,...) , x_j=0,1,
\end{equation}
is said to be a {\it collective} if there exist limits of relative
frequencies $\nu_N(0)=n(0)/N$ and $\nu_N(1)=n(1)/N$ where $n(\alpha),
\alpha=0,1,$ are numbers of realizations of the labels $\alpha$ in the
first $N$ trials.
\footnote{ Of ~course, in applications this means stabilization of digits
in the decimal expansion of relative frequencies}
The limits of ~these frequencies are probabilities in the framework of von
Mises frequency probability theory.

The main advantage of the Mises
approach with respect to the Kolmogorov one is that in the first one
there is some kind of an underground level before probabilities. This is
the level of collectives (random sequences). There are some situations
in physics (in particular, the two slit experiment) where we might not
compute all probabilities (at least, exactly), but we ~might extract
some properties of the corresponding random sequences using the ~known
probabilities.

The main line (a curve?)  of our ideas  is the following one.

The first thing which we know on the basis of the two slit experiments  is
that the corresponding random sequences are not Mises' collectives (because
probabilities have unusual properties). May
we generalize the Mises notion of the collective to get random sequences
which are more adequate to the two slit experiment? Yes, we can do this. The
most  general extension of Mises` theory is provided by the following
scheme [7]:

Let $\tau$ be an arbitrary topology on ${\bf Q}.$  The statistical
sample (ref{e1}) is said to be a $\tau$-{\it collective} if limits of relative
frequencies $\nu_N(\alpha),\alpha=0,1,$ exist with respect to the
topology $\tau.$ These limits belong to a completion ${\bf Q}_\tau$ of
${\bf Q}.$ The topology $\tau$ is said to be a topology of statistical
stabilization. For example, if $\tau$ is corresponding to a metric
$\rho$ on ${\bf Q}$ , then the $\tau$-stabilization means that
$\rho(\nu_N(\alpha),\nu_M(\alpha))\to 0 ,N,M\to\infty, \alpha=0,1.$ In
particular, if $\rho_R(x,y)=\vert x-y\vert_R$ is the ordinary real
metric, we obtain the old Mises theory.

The topology of statistical stabilization $\tau$ is a parameter of a
physical model. There are many physical experiments where $\tau$ is the
ordinary real topology. These experiments generate Mises collectives.
However, there are some experiments where topologies of statistical
stabilization are more exotic. We think that the two slit experiment
and other quantum interference experiments
belong to the last class of experiments.

The main (and very hard) problem is to find the right topology $\tau$ of
statistical stabilization corresponding to the fixed physical
experiment ${\cal E}.$ In many cases (especially in quantum
experiments)  this is really impossible to find $\tau$ exactly. However,
sometimes the known information about properties of statistical samples
generated by ${\cal E}$ gives us the possibility to describe a class
$C({\cal E})$ of possible topologies of statistical stabilization.
Then we can study theoretically the properties of $\tau$-collectives,
$\tau\in C({\cal E}),$ and obtain new properties which we cannot see
directly from a statistical data for ${\cal E}.$ Using these new
(theoretical) properties, we may try to explain some problems connected
with ${\cal E}.$ Of course, it would be only a theoretical explanation
To confirm it, we need to propose new physical experiments.
This was the main line of our ideas.

This scheme is realized in the present paper for the two slit
experiment. We shall show that if ${\cal E}={\cal E}_2$ is the two slit
experiment, the class $C_p({\cal E}_2)$ of $p$-adic topologies on
${\bf Q}$ seems to be adequate to this experiment. Then we shall study
the properties of generators of random numbers for $p$-adic topologies.
Using these properties, we could get some new predictions.

We notice that our $p$-adic topologies of statistical
stabilization are not so exotic. According to the famous theorem of
numbers theory (Ostrovsky theorem [17] ), every metric on ${\bf Q}$ of the
form $\rho(x,y)=\vert x - y\vert $ where $\vert\cdot\vert$ is an
absolute value (valuation) on ${\bf Q}$ coincides with the real one or
with one of  $p$-adic metrics.

\section{p-adic frequency realization of negative probabilities}

     The field of real numbers ${\bf R}$ is constructed
     as the completion of the field of rational numbers ${\bf Q}$
 with respect to
     the metric $\rho(x,y) =\vert x - y \vert$ , where $\vert \cdot\vert$ is
     the usual absolute value . The fields of $p$-adic
     numbers ${\bf Q}_p$
     are constructed in a corresponding way , by using other absolute values.
     For any prime number the $p$-adic absolute value
 $\vert \cdot\vert_p $ is defined in the
      following  way. At the first , we define it for natural numbers .
       Every natural number $n$ can be represented as the product of prime
       numbers : $ n = 2^{r_2}3^{r_3} \cdots p^{r_p} \cdots $. Then we define
       $\vert n\vert_p = p^{-r_p} $ , we set $\vert 0 \vert_p
       =0 $ and  $\vert -n\vert_p = \vert n \vert_p .$ We extend the definition of
$p$-adic absolute value $\vert\cdot\vert_p$ to all rational numbers by setting for
 $m\not=0 :$
       $ \vert n/m\vert_p= \vert n\vert_p/\vert m\vert_p. $ The completion
       of ${\bf Q}$ with respect to the metric $\rho_p (x,y)= \vert x- y\vert_p$
       is a locally compact field ${\bf Q}_p$ .
       It is well known , see [17 ] , that $\vert\cdot\vert$ and
$\vert\cdot\vert_p$ are the only possible absolute values on ${\bf Q}.$
The $p$-adic absolute value satisfies the strong triangle inequality :
$\vert x + y\vert_p\leq \max(\vert x\vert_p ,\vert y\vert_p) .$
For any $x\in Q_p$ we have a unique  canonical expansion ( converging in the
$\vert \cdot\vert_p$-norm ) of the form
\begin{equation}
\label{e2}
x= a_{-n}/p^n +\cdots\ a_{0}+\cdots+ a_k p^k+\cdots =
...a_k...a_0,a_{-1}...a_{-n},
\end{equation}
where $a_j=0,1,...,p-1 ,$ are the "digits" of the $p$-adic expansion.

Now we fix the prime number $p$ and choose the $p$-adic topology as the
topology of statistical stabilization , i.e. consider $p$-adic
collectives as random sequences. The following mathematical result [7] is
very important for our further considerations :

{\it Every $p$-adic number $x$ might be realized as a $p$-adic ~frequency
probability.}

For example,every rational number may be realized as a $p$-adic
probability. There are such ``pathological" probabilities (from the point
of view of the usual theory of probability) as ${\bf P}(A)=2 $ , ${\bf P}(A)=100$,
${\bf P}(A)=5/3$,${\bf P}(A)= -1;$
it may be possible that ${\bf P}(A)=i_p=\sqrt{-1} $ ,if $p=1 (mod 4),$
because in this case $i_p$ exists directly in $Q_p.$

The possibility to get negative probabilities using the frequency
definition is the most important motivation to choose the $p$-adic
topologies as topologies of statistical stabilization for the two slit
experiment, $C({\cal E}_2)=C_p({\cal E}_2).$

We continue the chain of our considerations. As we know from papers
of R.Feynman [3], all problems of a probabilistic description of the
two slit experiment might be solved on the basis of negative probability
distributions.\footnote{All Feynman`s investigations were heuristic,
because negative probabilities were meaningless from the mathematical
point of view.} As these probabilities may be realized as $p$-adic
frequency probabilities, we may assume that the two slit experiment
generates $p$-adic collectives (random sequences with statistical
stabilization in one of $p$-adic topologies).

{\bf Remark.} We are not sure that we have found the exact class of
topologies of statistical stabilization for the two slit experiment.
Probably $C_p$ is only the class of toy topologies which present only
one specific property $\pi_{ran}$ of random sequences generated by two slit
experiment. In the mathematical description this property is realized as
negative probabilities. The $p$-adic topologies present only this
particular property of random sequences in the two slit
experiment. We wish to find a physical counterpart $\pi_{phys}$
in the two slit experiment corresponding to the property $\pi_{ran}.$

\section{Algorithmic complexity of random sequences generated by the two
slit experiment}

The property $\pi_{ran}$ can be described on the basis of Kolmogorov`s
ideas about the algorithmic complexity of random sequences [18].
Kolmogorov's idea was  to define random sequences on the basis of a notion of a complexity of their
finite segments.
As usual, finite vectors $x=(x_1,...,x_n), x_j=0,1,$ are called words with
respect to the alphabet $\{0,1\}.$

{\bf Definition. } (A. N. Kolmogorov ) {\it  Let ${\cal A}$ be an arbitrary
algorithm. A complexity of a word $x$ with respect to ${\cal A}$ is
$$
K_{{\cal A}} (x) =\min l(\pi),
$$
where $\{\pi\}$ are the ~programs which are able to realize the word $x$
with the aid of ${\cal A}.$  }

This definition depends very much on a structure of ${\cal A}$. But
A.N.Kolmogorov proved the following theorem, which was a good justification
of this definition.

{\bf Theorem.}{\it There exists such algorithm ${\cal A}_0$ (optimal
algorithm) that
\begin{equation}
\label{e3}
K_{{\cal A}_0}(x) \preceq K_{{\cal A}}(x)
\end{equation}
for every algorithm ${\cal A}.$}
As usual, (ref{e3}) means that there exists such constant $C$ that $$
K_{{\cal A}_0}(x) \leq K_{{\cal A}}(x) + C $$ for all words $x$.
An optimal algorithm ${\cal A}_0$ is not unique.

{\bf Definition .}{\it The complexity ${\cal K}(x) $ of the word $x$ is
equal to the complexity $K_{{\cal A}_0}$ with respect to one fixed (for all
considerations) optimal algorithm ${\cal A}_0.$ }

A.N.Kolmogorov proposed to use the notion of the ~complexity of a finite
word to try to define a random sequences with the aid of complexities
of their finite segments. The idea of Kolmogorov  was very natural. He
proposed to consider a sequence $\omega\in \Omega $ as a random sequence,
if finite segments $(\omega)_n=(\omega_1,...,\omega_n) $ of this
sequence had complexities which are approximately equal to $n$.
Thus,a sequence $\omega$
is a random sequence in the Kolmogorov sense iff it is ~impossible to find
~programs $\pi_n$ ,generating words $(\omega)_n $ , with lengths $l(\pi_n)\ll
n.$  We need a word with a length not less then the length of the segment
of $\omega$ for coding this segment.
\footnote{In fact, the situation is not so simple from the mathematical point
of view.  We need to use more complicated notions of complexity.}

In [7] we estimated the Kolmogorov complexity of $p$-adic collectives.
The main result is that complexity of the initial segments of
a $p$-adic collective has the asymptotic $\log_p n.$ Hence the
Kolmogorov complexity for $p$-adic collectives is ~essentially less than
the same complexity for ordinary random sequences ( Mises` collectives).
Therefore previous trials contain some information about the next trial.
In particular, these trials are not independent. On the other hand,
correlations between trials are sufficiently weak since the complexity
${\cal K}((\omega)_n)$ increases as $log_p n.$ Thus we are not able to
predict a lot about the next trial on the basis of the previous results.

{\bf Concluding remarks.} 1).We wish again to notice that we are not sure that
statistical ~samples of the two slit experiment are $p$-adic collectives.
We cannot test this hypothesis directly (because we have not statistical
data for separate slits). Probably the corresponding topology of
statistical stabilization is much more complicated. We only extract one
property of this topology which has the adequate $p$-adic description.
As a consequence we obtain correlations between trials.

2) We note that $\log$-complexity is sufficiently high complexity. Therefore, it
 is natural that it could be
 identified with the linear complexity (which corresponds to ordinary
 random sequences) in some experiments. In fact, the experiments have to
 distinguish two following hypothesis:
 $ H_{Kol}= \{$ random sequences generated in interference experiments have the
 linear asymptotic of complexity $\}$ and
 $ H_p=\{ $ random sequences generated in interference experiments have the
 logarithmic asymptotic of complexity $\}.$  As we have already told,
 we could not be sure that
 the $\log$-behaviour is the right behaviour for ~interference experiments.
 Probably, $K((\omega)_n)=f(n),$ where $f(n)$ is some other function which
 increases slower that the linear function. Moreover, it should be that different ~interference experiments have different asymptotic behaviour, i.e.
 $f(n)=f_{\cal E}(n).$ In particular, it should be that
 $f(n)= \log_{p({\cal E})} n.$ It should be
 possible to classify interference ~phenomenon on this basis (2-adic interference
 and 1997-adic ~interference).

 3). The experiments to verify $\log$-correlations could be realized on the basis of the existed equipment for the neutron interferometry.
 The main problem is to attract physicists working in the neutron
 interferometry to realize these experiments. It is not easy, because there is
 the general opinion that the former experiments proved the independence of trials in the interference experiments.

I should like to thank L. Accardi, S. Albeverio, A. Holevo, G. Parisi
and  M. Namiki
for (sometimes critical) discussions on the possible applications of
$p$-adic probabilities in quantum theory. I should like to thank
H. Rauch,  J. Summhammer, M. Zawisky for (sometimes critical) discussions on
the possible experiments in the framework of the neutron interferometry
to test the $p$-adic predictions.

REFERENCES

[1] R. Feynman and A. Hibbs, Quantum mechanics and path integrals.
McGraw-Hill, New-York, 1965.

[2] A. N. Kolmogoroff,  Grundbegriffe der
Wahrscheinlichkeitsrechnung. Berl,1933. English translation by
N. Morrison, New-York, 1950.

[3] R. P. Feynman , Negative probability , in "Quantum Implications ",
Essays in Honour of David Bohm, B. J. Hiley and F. D. Peat, editors,
Routledge and Kegan Paul, London, 1987, 235 .

[4] G. Lochak, Found. of Physics {\bf 6}(1976), 173-184;

[5] G. Lochak, De Broglie initial conception of de Broiglie waves, 1-25 ;

L. Accardi, The probabilistic roots of the quantum mechanical
paradoxes, 297-330 .

The wave-particle dualism. A tribute to Louis de Broiglie on his 90th
Birthday, Edited by S. Diner, D. Fargue, G. Lochak and F. Selleri,
1970, D. Reidel Publ. Company, Dordrecht.

[6] W. Muckenheim , A review on extended probabilities,
 Phys. Reports, {\bf 133}  (1986) , 338-401 .

[7] A.Yu.Khrennikov ,$p$-adic valued distributions in mathematical
physic. Kluwer Academic Publishers, Dordrecht, 1994.

[8] V. S. Vladimirov, I. V. Volovich, E. I. Zelenov, $p$-adic numbers in
mathematical physics. World Sc. Publ., Singapure, 1993.

[9] T. L. Fine, Theories of probabilities, an examination of foundations.
 Academic Press, New-York (1973).

[10] K. R. Parthasarathy, An introduction to quantum stochastic calculus.
Birkhauser, Basel, 1992.

[ 11] H. Rauch , J.
Summhammer, M. Zawisky, E. Jericha, Law-contrast
and low-counting-rate measurements in neutron interferometry. Phys. Rev. A,
{\bf 42}, 3726-3732 (1990).

[12] R. von Mises, Probability, Statistics and Truth.
 Macmillan, London, 1957.

[13] V. Buonomano, Quantum uncertainties, Recent and Future Experiments and Interpretations , edited by W.M. Honig, D.W. Kraft and E. Panarella,
NATO ASI Series, {\bf 162}, Plenum Press, New York (1986).

[14]  M. Zawisky, H. Rauch, Y. Hasegawa,
Contrast enhancement by time selection in neutron interferometry. {\it Phys. Rev. A}, {\bf 50}, 5000-5006
(1994).

[15] J. Summhammer, Neutron interferometric test of the nonergodic interpretation of quantum mechanics. {\it Il Nuovo Cimento}, {\bf 103 B}, 265-280 (1989).

[16] M. Zawisky, {\it Zeitselektive neutronen-interferometrie.}
Dissertation, Technischen Univ. Wien (1993).

[17] W. Schikhof, Ultrametric Calculus. Cambridge Studies in
 Adv. Math. 4.Cambridge U.P.Cambridge (1984).

[18] A. N. Kolmolgorov, Logical basis for information theory and probability
theory. IEEE Trans. {\bf IT-14}, 662-664 (1968).

[19] A. Yu. Khrennikov, On the experiments to find $p$-adic stochastic in the two slit
experiment, Preprint Ruhr-University Bochum, SFB - 237, No. 309 (1996).

\end{document}